# Organ Dose Equivalents of Albedo Protons and Neutrons Under Exposure to Large Solar Particle Events during Lunar Human Landing Missions

Sungmin Pak and Francis A. Cucinotta*

*Department of Health Physics and Diagnostic Sciences, School of Integrated Health Sciences,*

*University of Nevada, Las Vegas, NV 89154, USA*

*Correspondence author

E-mail:  francis.cucinotta@unlv.edu

*Correspondence to:

Professor Francis A. Cucinotta,

University of Nevada Las Vegas,

Department of Health Physics and Diagnostic Sciences

Las Vegas, NV, 89154, USA

Tel:     +1 702-895-0977

Fax:     +1 702-895-1353

E-mail:  francis.cucinotta@unlv.edu

**Abstract**

Astronauts participating in lunar landing missions will encounter exposure to albedo particles emitted from the lunar surface as well as primary high-energy particles in the spectra of galactic cosmic rays (GCRs) and solar particle events (SPEs). While existing studies have examined particle energy spectra and absorbed doses in limited radiation exposure scenarios on and near the Moon, comprehensive research encompassing various shielding amounts and large SPEs on the lunar surface remains lacking. Additionally, detailed organ dose equivalents of albedo particles in a human model on the lunar surface have yet to be investigated. This work assesses the organ dose equivalents of albedo neutrons and protons during historically large SPEs in August 1972 and September 1989 utilizing realistic computational anthropomorphic human phantom for the first time. Dosimetric quantities within human organs have been evaluated based on the PHITS Monte Carlo simulation results and quality factors of the state-of-the-art NASA Space Cancer Risk (NSCR) model, as well as ICRP publications. The results with the NSCR model indicate that the albedo contribution to organ dose equivalent is less than 3% for 1 g/cm$^2$ aluminum shielding, while it increases to more than 20% in some organs for 20 g/cm$^2$ aluminum shielding during exposure to low-energy-proton-rich SPEs.

# 1. Introduction

Space radiation poses significant health risks for crewed space travel (Cucinotta, 1999; Cucinotta & Durante, 2006; Durante & Cucinotta, 2011; Cucinotta, 2022). Unlike on Earth's surface, where the magnetosphere and atmosphere offer substantial protection against space radiation, astronauts undertaking lunar landing missions will face a harsh radiation environment with limited natural shielding (Cucinotta et al., 2010). The primary sources of space radiation beyond low Earth orbit (LEO) are galactic cosmic rays (GCRs), comprising high-energy protons and high-energy and charge (HZE) nuclei, and solar particle events (SPEs), predominantly consisting of low- to medium-energy protons.

Space radiation characteristics on and near the Earth's Moon have been studied by measurements and simulations. The Cosmic Ray Telescope for the Effects of Radiation (CRaTER) aboard the Lunar Reconnaissance Orbiter (LRO) and the Lunar Lander Neutrons and Dosimetry (LND) dosimeter aboard the Chang'E 4 lander indicates an absorbed dose of 13.6 cGy/yr behind 3 g/cm$^2$ aluminum-equivalent shielding at 50 km above the lunar surface (Spence et al., 2013) and 11.6 cGy/yr behind 1 g/cm$^2$ aluminum-equivalent shielding on the lunar surface (Zhang et al., 2020). Simulation studies (Spence et al., 2013; Zaman et al., 2020) have suggested an approximately 8 to 9% of albedo ion, neutron, electron, and positron contribution to the absorbed dose, and a skin dose equivalent of 105 cSv/yr for GCR exposure behind thin aluminum shielding based on quality factors (QF) specified in the International Commission on Radiological Protection (ICRP) Publication 60 (ICRP, 1991).

While the previous research and measurements offer valuable insight into the lunar radiation environment, the health effects of primary and albedo particles on various human organs under different shielding amounts have not been thoroughly explored for large SPEs, which are a critical potential concern for lunar human landing missions. Satellite data indicates that more than 70 SPEs are typically recorded in one solar cycle, with over 20% of these featuring protons of 30 MeV or higher energy (Shea & Smart, 1990; Kim et al., 2009). While many SPEs result in negligible radiation dose due to their low flux and energy, notably intense solar storms, such as those observed in August 1972 and series of events in August, September, and October of 1989, can lead to significantly high doses for light to medium shielding amounts (<20 g/cm$^2$) (Kim et al., 2009; Jiggens et al., 2014; Cucinotta & Pak, 2024).

This study investigates the organ dose equivalents in a realistic human male phantom behind various aluminum shielding thicknesses on the lunar surface during exposure to historically large SPEs in August 1972 and September 1989. The August of 1972 event has one of the largest proton numbers at low to medium energies (<100 MeV), while the September of 1989 one of the largest events at high energies (>100 MeV). Our investigation estimates the contribution of albedo protons and albedo neutrons to the organ dose equivalents. Dosimetric quantities within human organs have been assessed using the latest NASA Space Cancer Risk model (NSCR-2022) (Cucinotta et al., 2017; Cucinotta, 2024), as well as references from ICRP (ICRP, 1991; ICRP, 2007) for comparative analysis.

## 2. Methods

To comprehensively evaluate organ dose equivalents behind diverse shielding amounts on the lunar surface, a series of simulation works have been conducted utilizing the PHITS3.27 simulation toolkit, which is designed specifically for heavy ion transport in space research and accelerator studies (Iwase, Niita, & Nakamura, 2002; Niita et al., 2006; Iwamoto et al., 2022; Sato et al., 2024).

To consider practical radiation exposure scenarios on the lunar surface with various shielding thicknesses, ranging from 1 g/cm$^2$ of a space vehicle to 20 g/cm$^2$ thicknesses of typical spacecraft, a hemispherical structure of aluminum with thicknesses of 1, 2, 10, and 20 g/cm$^2$ has been reconstructed on the lunar surface. Lunar soil in the ground has a density of 1.5 g/cm$^2$ and consists of 43.67% oxygen, 0.32% sodium, 5.56% magnesium, 9.00% aluminum, 21.18% silicon, 8.49% calcium, 1.46% titanium, 0.13% manganese, and 10.19% iron, based on lunar soil samples (Prettyman et al., 2006; Zaman et al., 2020).

To ascertain dosimetric quantities in human crews for lunar landing missions, the three-dimensional reference adult male voxel phantom introduced in ICRP Publication 110 (ICRP, 2009) has been reconstructed within the hemispherical radiation shielding structure on the lunar surface.

Simulations have been conducted separately for primary SPE spectra and for the upward albedo protons and neutrons during each SPE exposure scenario. Primary SPE spectra for significant historical events in August 1972 and September 1989 have been generated based on the Tylka model (Tylka & Dietrich, 2009; Tylka, Dietrich, & Atwell, 2010), which expresses integral SPE proton fluence using the Band function (**Table 1** and **Fig. 1**). Energy spectra of the upward albedo protons and neutrons for each SPE exposure scenario have been obtained at 1 micrometer above the lunar surface using the [T-Cross] tally and then re-generated in separate codes to investigate the dosimetric quantities of albedo particles only. For precise space radiation transport, the JQMD-2.0 physics model has been adopted, which is the advanced Quantum Molecular Dynamics (QMD) model developed by the Japan Atomic Energy Research Institute (JAERI) to simulate accurate hadronic collisions and provide reliable data on secondary neutron and heavy fragment production (Niita et al., 1995; Ogawa et al., 2015a; Ogawa et al., 2015b; Ogawa et al., 2016).

A series of simulations of primary SPE spectra and albedo protons and neutrons provided absorbed doses and fluences of protons, deuterons, tritons, neutrons, pions, kaons, muons, electrons, positrons, and ions with Z=2-28 in critical human organs for each scenario with different shielding amounts.

Simulation results have been converted into organ dose equivalent ($H$ [Sv]) by:

$$H = D \cdot QF \qquad (1)$$

where $D$ [Gy] is the absorbed dose, and $QF$ stands for the radiation quality factor.

The ICRP quality factor, which was developed for ground-based exposures, is defined by (ICRP, 1991; ICRP, 2007):

$$QF(L) = \begin{cases} 1, & L < 10 \ keV/\mu m \\ 0.32L - 2.2, & 10 \ keV/\mu m \leq L \leq 100 \ keV/\mu m \\ 300/\sqrt{L}, & L > 100 \ keV/\mu m \end{cases} \quad (2)$$

where $L$ [keV/μm] is the linear energy transfer (LET).

In the NSCR model, considering the microscopic energy deposition or track structure (Cucinotta et al., 2017; Cucinotta, 2024) is considered to develop a space radiation QF. For largely low LET protons the effects of non-targeted effects (Cucinotta, 2024) are not considered and the QF is defined by:

$$QF(Z,E) = Q_L(Z,E) + Q_H(Z,E) \quad (3)$$

where

$$Q_L(Z,E) = [1 - P(Z,E)] \quad (4)$$

$$Q_H(Z,E) = 6.24\Sigma_0 P(Z,E)/(\alpha_\gamma L) \quad (5)$$

$$P(Z,E) = [1 - \exp(-Z^{*2}/\kappa\beta^2)]^m [1 - \exp(-E/0.2)] \quad (6)$$

Here, $E$ is the particle's kinetic energy in the unit of [MeV/u] for nuclei and [MeV] for other particles, $L$ [keV/μm] is the particle's LET, $Z$ is the particle's charge number, $Z^* = Z[1 - \exp(-125\beta/Z^{2/3})]$ is the particle's effective charge number, and $\beta$ is the particle's relative speed to the light. The other parameters used in this calculation are listed in **Table 2**. The NSCR model accounts for secondary electrons with sufficient energy to cause additional ionizations nearby, called delta-rays, of which generation is proportional to $Z^{*2}/\beta^2$.

## 3. Results

The results for the simulations of primary SPE spectra are the summation of the dosimetric quantities of primary and albedo particles. Conversely, the simulations of upward albedo particle spectra provide the dosimetric quantities of albedo particles only, leading to the assessment of albedo particle contribution to the organ dosimetric quantities by comparing with the results for primary SPE simulations. Particle-specific and organ-specific dose equivalents during exposure to two large SPEs, as well as the contribution of albedo neutrons and protons, are elucidated in this section.

**Figs. 2 and 3** depict organ dose equivalents behind various aluminum shielding depths for SPE in August 1972, assessed with the NSCR and ICRP models, respectively, while **Figs. 4 and 5** show the results for SPE in September 1989. It has been demonstrated that dose equivalent during exposure SPEs drastically decreases with increasing shielding. In addition, a significant difference in dose equivalent from organ to organ is indicated with high dose equivalent in external organs, such as skin and breast, and low dose equivalent in internal organs, such as bladder and colon. Because the 1972 event has higher total proton fluence with a massive amount of protons with E < 100 MeV, the dose equivalent is higher for the 1972 event compared to the 1989 event with thin (1 and 2 $g/cm^2$) shielding, while it is opposite for thicker (10 and 20 $g/cm^2$) shielding due to increased number of protons with higher energy ($\geq$ 100 MeV) in the 1989 event spectra. While most of the secondary particles generated for the 1972 event are neutrons, followed by protons and alpha particles, other secondary ions and mesons also take place, as well as neutrons and alpha particles for the 1989 event. Compared to the ICRP model, the NSCR model suggests higher proton and meson dose equivalents and lower heavy ion dose equivalents. Since the majority of dose equivalent is induced by protons for SPE exposure, the NSCR model estimates a higher total dose equivalent compared to the ICRP model.

**Figs. 6-9** demonstrate the contributions of albedo neutrons and protons to NSCR and ICRP organ dose equivalents for SPEs in August 1972 and September 1989. The proportion of albedo dose equivalents in total dose equivalents significantly rises with shielding depth. Assuming a human in a standing position, the contribution of albedo particles is more considerable in internal pelvic organs (bladder and prostate) compared to the head and neck (brain, salivary glands, and thyroid) for thin aluminum shielding. This is because the dose equivalent from primary particles is minimal in the bladder and prostate due to a large

amount of tissue shielding, and the organs in the lower body receive a larger dose from albedo particles compared to the organs in the upper body. On the other hand, as the amount of aluminum shielding increases, the albedo contribution to external organs (skin and gonads) drastically increases due to the dose equivalent from primary particles significantly decreasing in such organs. It has also been indicated that the albedo protons play a minor role specifically in internal organs for low-energy-proton-rich SPE in August 1972. Conversely, they make a higher contribution in dose equivalent in most organs for high-energy-proton-rich SPE in September 1989, while albedo neutrons dominate. The results also suggest that the albedo contribution is more significant for the 1972 event compared to the 1989 event, especially for heavy aluminum shielding because of decreased dose equivalent from the primary particles and increased low-energy albedo neutron generation in the ground. The NSCR and ICRP models suggest similar albedo contributions, while the ICRP model indicates slightly higher values since it estimates lower primary dose equivalents than the NSCR model.

## 4. Discussion

Organ dose equivalents and the contribution of albedo neutrons and protons in critical human organs have been assessed for exposure to historically large SPEs on the lunar surface. For thin shielding, albedo contribution to dose quantities for SPE exposure is suggested to be smaller than the measurement data for GCR exposure, while a dramatic increase in the contribution of SPE albedo particles to organ dose equivalent is indicated for thick shielding. The albedo contribution is notable for the event with extreme total proton flux, such as SPE in August 1972, compared to the high-energy-proton-rich SPE in September 1989.

Electron and positron dose quantities are shown to be ignorable in the results of this work, while high albedo electron and positron fluxes have been suggested by previous studies (Looper et al., 2013; Zaman et al., 2022). PHITS provides the Electron-Gamma Shower 5 (EGS5) algorithm (Hirayama et al., 2005) for more detailed electron, positron, and photon interactions in matter, while the Evaluated Photon Data Library 1997 (EPDL97) (Cullen, Hubbell, & Kissel, 1997) is adopted in this work due to the limitations of computing resources. Utilization of the EGS5 algorithm in future work may result in increased estimates of electron and positron dose quantities on the lunar surface.

The major components of uncertainties are the representation of the proton energy spectra, transport code including nuclear cross sections, radiation quality factors, and human geometry model. In the present report we use the re-analysis of proton energy spectra made by Tylka and Dietrich (2009). Because SPE spectra are highly variable (Kim et al, 2009) the present calculations provide a benchmark for specific spectra, however do not address the uncertainties in spectra. The current study has used the PHITS code and the ICRP recommended human phantom model (ICRP, 2009). The PHITS model uncertainties are largely in the cross-section models used in calculations. However, statistical error also occurs in Monte-Carlo calculations. These are found to be relatively small in large organs, such as the skin and liver, behind thin shielding is mostly less than 3%, it increases to more than 20% for small organs, such as the prostate and thyroid, behind thick shielding.

The NSCR model is unique in its assignment of uncertainty distributions in the QF parameters (Cucinotta et al., 2017; Cucinotta 2024), while uncertainties are not considered using the ICRP defined QF. The QF uncertainties are combined with other contributors to cancer and circulatory disease risks predictions in the NSCR model (Cucinotta, 2024) to estimate an overall uncertainty in risk predictions.

The choice of the human geometry model can play an important role in organ dose evaluations. Previous estimates by Kim et al. (2010) comparing the CAM model based on a combinatorial geometry approach to the MAX/FAX model based on CT scan data of humans showed important differences for SPEs with only minor differences for GCR. This is due to the steep dose gradients that results for many SPE spectra. These differences will be larger for organs with less self-shielding such as the skin (Kim et al., 2006) and lens (Cucinotta et al., 2001). The present calculations use the ICRP recommended phantom model for average males. For the steep dose gradients of SPE spectra differences based on size and weight and between males and females will occur. In addition, further investigation into positions other than standing and simulations for female crews is necessary for a comprehensive understanding of the health issues of SPE albedo particles during lunar human landing missions. These variations can be considered in future work.

## Acknowledgments

No funding was received for this study. The simulation works were performed on the Cherry-Creek Cluster of the National Supercomputing Institute (NSI) at the University of Nevada, Las Vegas (UNLV).

## Conflict of Interest

The authors declare no conflicts of interest.

**Table 1**
Band function parameters for SPEs on August 4, 1972, and September 29,1989 (Tylka & Dietrich, 2009; Tylka, Dietrich, & Atwell, 2010).

| Event | | $J_0$ [protons/cm$^2$] | $\gamma_1$ | $\gamma_2$ | $R_0$ [GV] |
|---|---|---|---|---|---|
| August 4, 1972 | | $1.450 \times 10^{15}$ | -3.636 | 7.95 | 0.0345 |
| September 29, 1989 | (first 75 minutes) | $1.799 \times 10^5$ | 2.060 | 2.63 | 3.6593 |
| | (next 61 hours) | $2.027 \times 10^{10}$ | -0.109 | 4.58 | 0.0945 |

**Table 2**
Parameters in the NSCR model (Pak & Cucinotta, 2024).

| Parameter | Low Z (Z ≤ 2) | High Z (Z > 2) |
|---|---|---|
| $m$ | $3 \pm 0.5$ | $3 \pm 0.5$ |
| $\kappa$ | $624 \pm 69$ | $1000 \pm 150$ |
| $\Sigma_0/\alpha_\gamma$ [μm² Gy] | $(4728 \pm 1378)/6.24$ for solid cancer<br>$(1750 \pm 250)/6.24$ for leukemia | $(4728 \pm 1378)/6.24$ for solid cancer<br>$(1750 \pm 250)/6.24$ for leukemia |

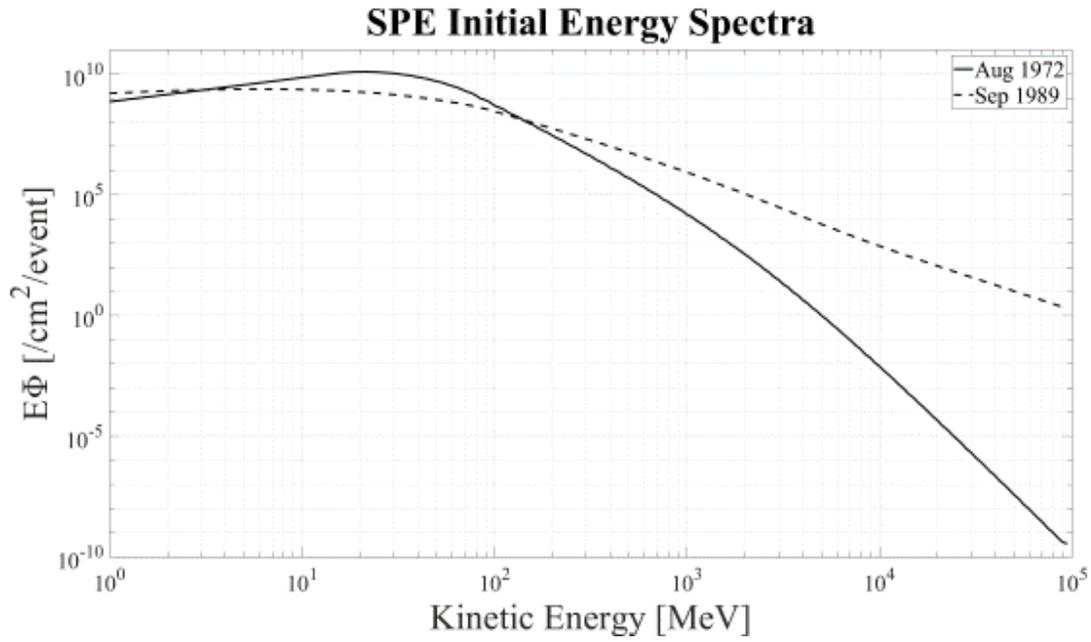

**Fig. 1.** Initial proton spectra generated in simulations to reconstruct SPEs in August 1972 and September 1989.

**Fig. 2.** Organ dose equivalent behind aluminum shielding with thicknesses of 1, 2, 10, and 20 g/cm$^2$ for SPE in August 1972 on the lunar surface, assessed with the NSCR model.

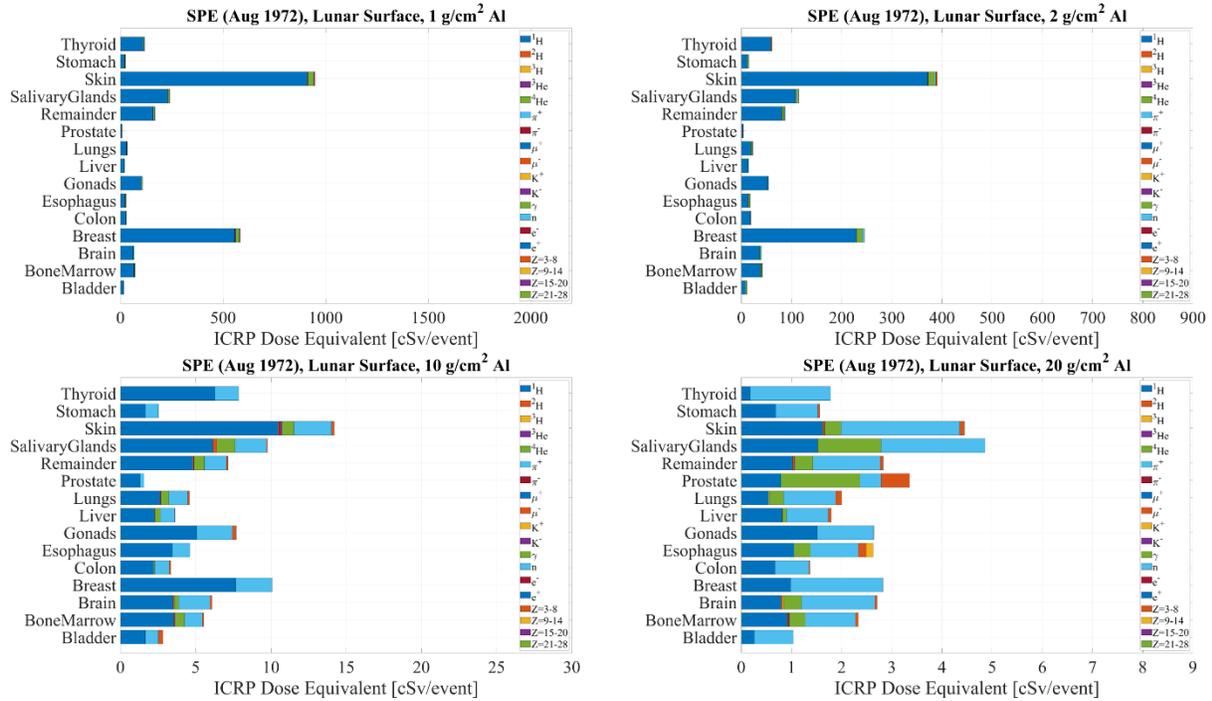

**Fig. 3.** Organ dose equivalent behind aluminum shielding with thicknesses of 1, 2, 10, and 20 g/cm$^2$ for SPE in August 1972 on the lunar surface, assessed with the ICRP model.

**Fig. 4.** Organ dose equivalent behind aluminum shielding with thicknesses of 1, 2, 10, and 20 g/cm$^2$ for SPE in September 1989 on the lunar surface, assessed with the NSCR model.

**Fig. 5.** Organ dose equivalent behind aluminum shielding with thicknesses of 1, 2, 10, and 20 g/cm$^2$ for SPE in September 1989 on the lunar surface, assessed with the ICRP model.

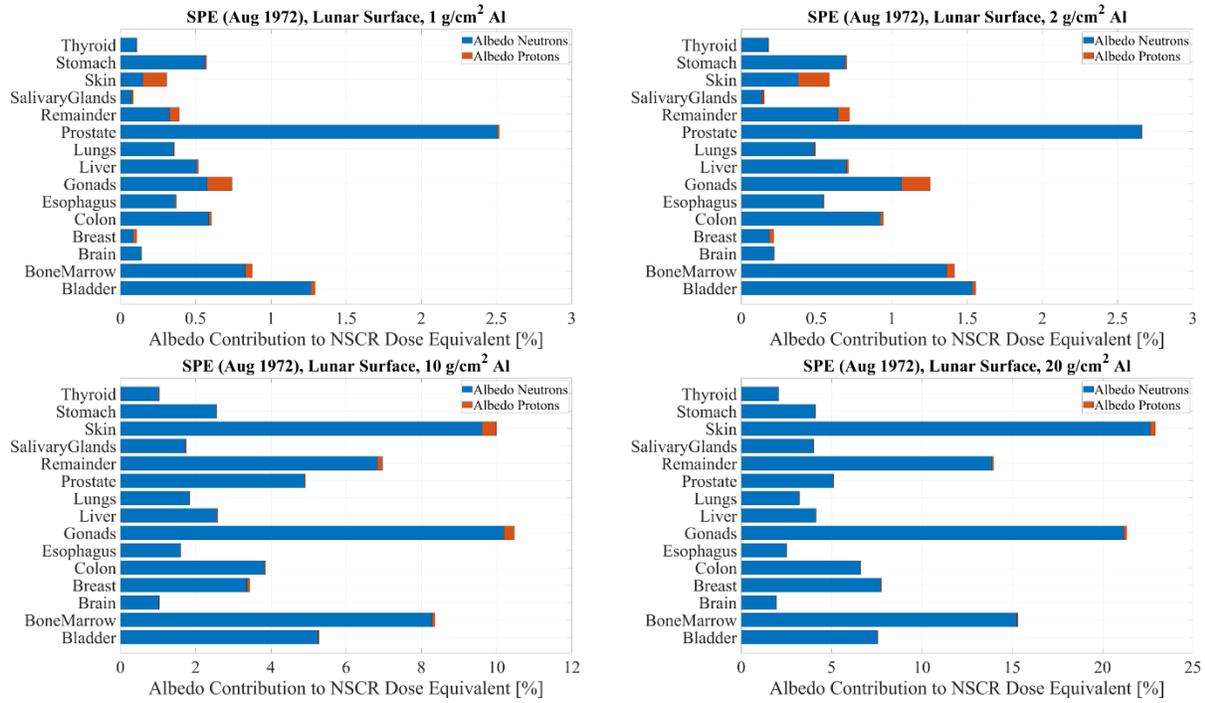

**Fig. 6.** Contribution of albedo neutrons and albedo protons to the NSCR dose equivalent behind various aluminum shielding thicknesses for SPE in August 1972 on the lunar surface.

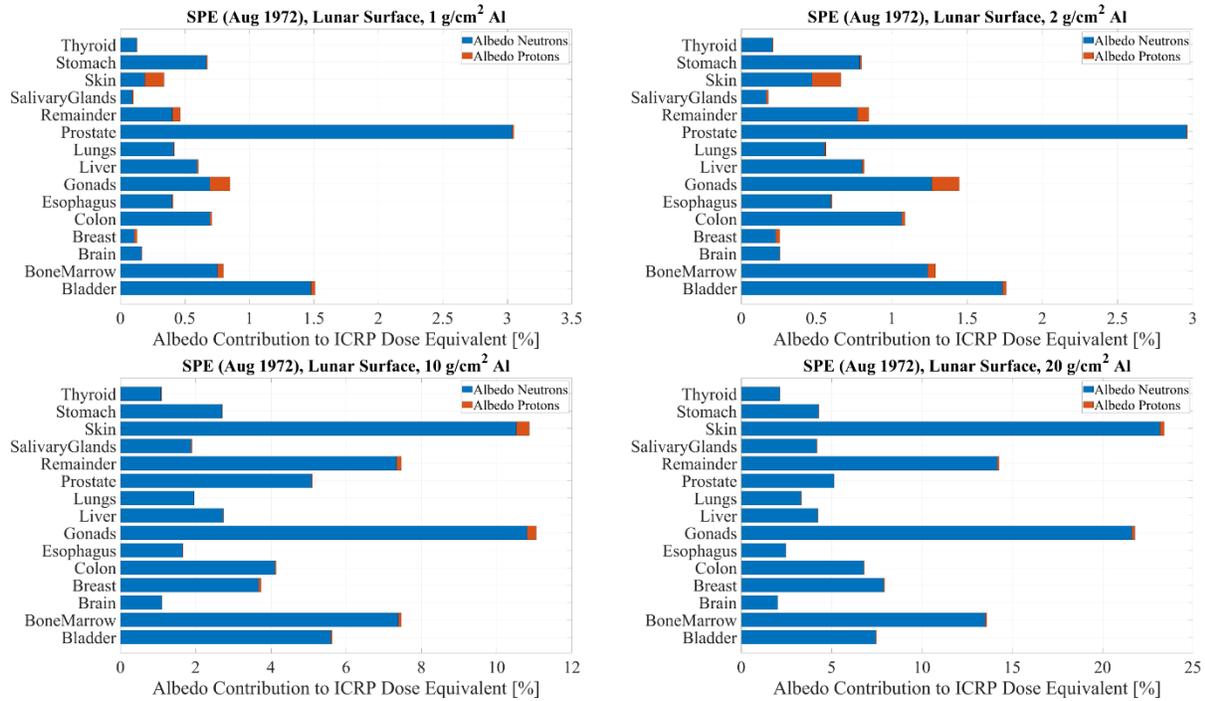

**Fig. 7.** Contribution of albedo neutrons and albedo protons to the ICRP dose equivalent behind various aluminum shielding thicknesses for SPE in August 1972 on the lunar surface.

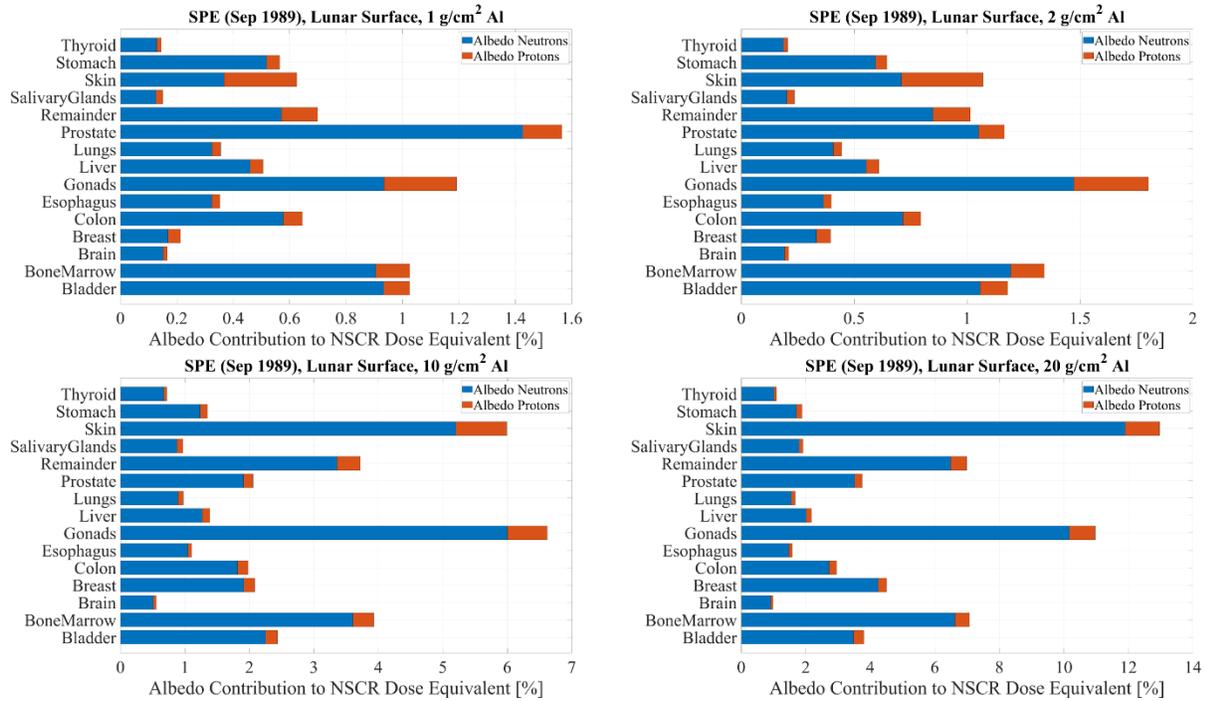

**Fig. 8.** Contribution of albedo neutrons and albedo protons to the NSCR dose equivalent behind various aluminum shielding thicknesses for SPE in September 1989 on the lunar surface.

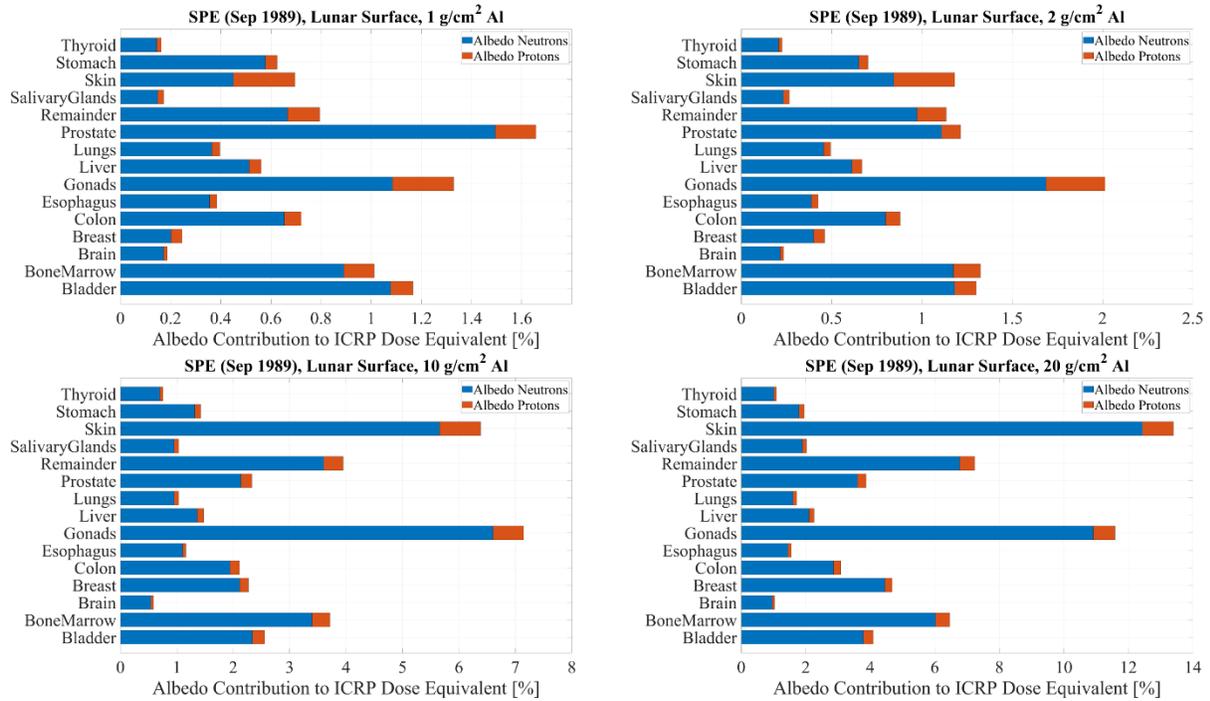

**Fig. 9.** Contribution of albedo neutrons and albedo protons to the ICRP dose equivalent behind various aluminum shielding thicknesses for SPE in September 1989 on the lunar surface.